\documentclass[prb,twocolumn,superscriptaddress,showpacs]{revtex4}

\def\beq{\begin{equation}}
\def\eeq{\end{equation}}
\def\beqn{\begin{eqnarray}}
\def\eeqn{\end{eqnarray}}

\newcommand{\ket}[1]{|#1\rangle}
\newcommand{\bra}[1]{\langle #1|}

\usepackage{graphicx}
\begin{document}
\title{Conservation and persistence of 
spin currents and their relation to 
the Lieb-Schulz-Mattis twist
operators}

\author{N. Bray-Ali}
\affiliation{Department of Physics and Astronomy, University of Southern California, Los Angeles, CA 90089, USA}
\author{Z. Nussinov}
\affiliation{Department of Physics, Washington University, MO 63160 USA}

\date{\today}

\begin{abstract}
Systems with spin-orbit coupling do not conserve ``bare'' spin current $\bf{j}$.  A recent proposal for a conserved spin current $\bf{J}$ [J.~Shi {\it et.al} Phys. Rev. Lett. {\bf 96}, 076604 (2006)] does not flow persistently in equilibrium.  We suggest another conserved spin current $\overline{\bf{J}}$ that may flow persistently in equilibrium.  We give two arguments for the instability of persistent current of the form $\bf{J}$: one based on the equations of motions and another based on a variational construction using Lieb-Schulz-Mattis twist operators.  In the absence of spin-orbit coupling, the three forms of spin current coincide.            
\end{abstract}
\maketitle

\section{Introduction}

Spintronics aims to manipulate the electron's
spin instead of its charge. 
\cite{spintronics1,spintronics2, Wolf,Zutic} 
A requisite for spintronic devices is the
generation of a spin current. \cite{Stevens,Habner} 
In recent years, exciting new proposals and predictions have
been made regarding the possibility of dissipationless
spin currents- primarily through the use of
the Spin Hall Effect (SHE). \cite{diss1,diss2} The 
SHE may generate sizable Hall spin currents and open the 
door for the realization of many of these notions. 
Observation of the SHE have been reported by
employing the Kerr effect in n-doped GaAs \cite{awschalom} and 
a p-n junction light emitting diode system. \cite{sinova}
Much fundamental work on spin currents has been carried out,
e.g., Ref.~\onlinecite{Fie,Mis,Wan,sun07,sun08,sonin}. 

A fundamental problem in spintronics is the non-conservation of the bare spin
currents (the sum of the product of each electron spin multiplied
by its velocity, see Eq.~(\ref{js})). This non-conservation
is triggered by 
spin-orbit effects. Charge conservation and the continuity equations
that it gives rise to are of crucial importance in electronic circuits.
Is there an analogue for spins? The quest for conserved spin 
currents has been the 
subject of much attention.\cite{niu,yang,jan}
An early step was made in Ref.~\onlinecite{niu}
that proposed a conserved spin current by augmenting the 
bare spin currents by an additional contribution 
so as to include the effects of torques triggered by spin-orbit effects. 
The work of Ref.~\onlinecite{niu} assumed that 
the average torque vanishes in the bulk and that the ``torque dipole
density'' on the system's surface can also be neglected.  The ``torque dipole
density'' was a field whose divergence led to the 
actual torque acting on spins at a particular location. 
This proposal was exciting. The spin accumulation observations
of Ref.~\onlinecite{awschalom} and Ref.~\onlinecite{sinova} 
indeed cannot measure the bare spin current.

We suggest that the notions of conservation and of persistence are independent for spin currents.  For example, the bare spin current in Eq.~(\ref{js}) in general is not conserved, and may persist in equilibrium (See Eq.~(\ref{central_js})).  Rashba noticed the possibility of persistent equilibrium flow of the bare spin current for a particular class of spin-orbit coupled models.\cite{rashba}  Here we present a model-independent argument for the result.  In contrast to the bare spin current, the conserved spin current ${\bf J}$ given by Eq.~(\ref{shi}) does not flow in energy eigenstates so any persistent flow implies the system is not in equilibrium.  Furthermore, we have found another conserved spin current $\overline{\bf{J}}$ and associated spin density $\overline{\bf{\rho}}$ that may flow persistently.  When spin is a good quantum number, these distinctions collapse: the bare spin current is conserved, and none of the three spin currents flow persistently.  However, in a generic, spin-orbit coupled system, the definitions of spin current differ with respect to their conservation and persistence.  

In Section (\ref{A}), we recall the standard definition of spin density and bare spin current.  We then review the conserved spin current proposed in Ref.~\onlinecite{niu}, and contrast it with an alternative, conserved spin current (See Eq.~(\ref{spdensity})).  Section (\ref{B}) further distinguishes this novel, conserved current by computing its expectation in a general spin-orbit coupled system.  We show that, in contrast to the previous proposal, it may flow persistently, just like the bare spin current.  In Section (\ref{A+}) we give a physical argument to explain this distinction: any state that carries the conserved spin current proposed by Ref.~\onlinecite{niu} may dissipate energy by relaxing to a lower energy state, which we explicitly construct.  The argument extends the one used by Bohm to establish that persistent charge current flows only in meta-stable states, not the ground-state.\cite{bohm}  Further, the construction generalizes one used by Lieb et al. to study quasi-one-dimensional, spin-1/2 systems.\cite{lsm}  We conclude with a summary and discussion of our results.

\section{Definitions of spin current and spin density}
\label{A}

Let us consider a general, interacting system with Hamiltonian
$H[\{\bf{r}_{i}\},\{\bf{p}_{i}\}, \{\sigma_{\alpha}\}]$. 
Here, $\{\bf{r}_{i}\}$ and $\{\bf{p}_{i}\}$
denote the phase space variables and  
$\{\sigma_{\alpha}\}$ reflect the spin states of
all non-scalar particles in the system.  The spin density along a given quantization axis (taken to be the $z-$axis) has the form 
\begin{eqnarray}
\rho =  \sum_{i} \sigma_{i}^{z} n_{i}, 
\label{density}
\end{eqnarray}
where, $n_i$ is the particle density, and $\sigma_i^{z}$ is the spin operator along the $z-$axis, given by a Pauli matrix for spin one-half particles.  In Ref.\onlinecite{niu}, Shi et. al. suggested the following form for the spin current:
\begin{eqnarray}
{\bf{J}} &=& \frac{d}{dt} \Big( 
\sum_{i} \sigma_{i}^{z} {\bf{x}}_{i} \Big) \nonumber \\
&=& \bf{j}+ \sum_{i} \tau_i^{z} {\bf {x}}_{i},
\label{shi}
\end{eqnarray}
where, the spin torque is given by $\tau_i^{z}=d\sigma_{i}^{z}/dt=1/(i\hbar)[\sigma_i^z,H]$.  In the absence of spin-orbit coupling, the spin torque vanishes, and the spin current $\bf{J}$ coincides with the ``bare'' spin current
\begin{eqnarray}
\bf{j} = \sum_{i} \sigma_i^z \bf{v}_i,
\label{js}
\end{eqnarray}
where, the velocity is given by ${\bf v}_{i}=d{\bf x}_i/dt=1/(i\hbar)[{\bf x}_{i},H].$

In contrast to the bare spin current (\ref{js}), the current in (\ref{shi}) satisfies the continuity equation, even in systems with spin-orbit coupling:\cite{niu}
\begin{eqnarray}
\frac{\partial \rho}{\partial t}  
+ {\bf \nabla} \cdot {\bf J} =0.
\label{thatsit}
\end{eqnarray}
Physically this means that in a system with boundary, the flow of spin current (\ref{shi}) across the boundary into the system leads to spin accumulation within the system.  We note that conservation of ${\bf{J}}$ only occurs when the average torque vanishes: $\sum_i \langle \tau_i^{z}\rangle =0$.\cite{niu}         

Interestingly, it is possible to construct another spin current $\overline{\bf{J}}$ and spin density $\overline{\rho}$ that satisfy the continuity equation.  Consider the following spin current:
\begin{eqnarray}
\overline{{\bf{J}} } &=& \sum_{i} \sigma_{i}^{z}(t=0) {\bf v}_{i} 
 \nonumber \\
 &=& {\bf j} - \sum_{i} \tau_i^{z} {\bf x}_{i} + \ldots,
\label{spdensity}
\end{eqnarray}
where, the omitted terms are higher order in the spin-orbit coupling.  We define a corresponding spin density $\overline{\rho} = \sum_{i} \sigma_{i}^{z}(t=0) n_{i}$ by evaluating at the same reference time.  Remarkably, this form of the spin density and spin current also satisfies the continuity equation:
\begin{eqnarray}
\frac{\partial \overline{\rho}}{\partial t}  
+ {\bf \nabla} \cdot \overline{{\bf J}} =0.
\label{overline_conserve}
\end{eqnarray}
To see this, notice that the label $\sigma_{i}^{z}(t=0)$ does not evolve, but merely labels each particle.  From the continuity of charge, the continuity of these labels follows.  We remark that the current $\overline{\bf{J}}$ obeys the continuity equation (\ref{overline_conserve}) even when the average torque does not vanish, in contrast to the current $\bf{J}$.  In the appendix, we perform the expansion in spin-orbit coupling explicitly.    
 
\section{Constraints on spin current from equations of motion}
\label{B}
As noted in Ref.~\onlinecite{niu}, the form of the spin current in Eq.~(\ref{shi}) does not flow in energy eigenstates.  We give a simplified derivation, then extend it to finite temperature, and use the result to deduce constraints on equilibrium flow of the other two forms of spin current Eq.~(\ref{js}) and Eq.~(\ref{spdensity}).  

The spin displacement operator
\begin{eqnarray}
{\bf L} = \sum_{i} {\bf x}_{i} \sigma^{z}_{i}
\label{aspin}
\end{eqnarray}
allows us to evaluate the spin current in Eq.~(\ref{shi}):
\begin{eqnarray}
\langle n| {\bf J }| n \rangle = \langle n| \frac{  d{\bf L}}{dt}| n \rangle 
=  - i  \langle n| [ {\bf{L}}, H] | n \rangle =0.
\label{bigeasy}
\end{eqnarray}
Here, $|n \rangle$ denotes the $n$th eigenstate of $H$ with energy $E_{n}$. 
The second equality in Eq.(\ref{bigeasy}) follows from the equations of motion
while the third follows as $\langle n|$ and $| n \rangle$ refer to
the same eigenstate of $H$ with specific eigenvalue ($E_{n}$).
Here and henceforth, we set $\hbar =1$. 

As an immediate corollary, we notice that the spin current in Eq.~(\ref{shi}) vanishes at finite temperatures as well:
\begin{eqnarray}
\langle {\bf{J}} \rangle =  \frac{\sum_{n} \langle n| {\bf{J}}| n \rangle e^{-\beta E_{n}}}
{\sum_{n} e^{-\beta E_{n}}} = 0,
\label{equil}
\end{eqnarray}
where, the  inverse temperature is $\beta=1/k_BT$. 

If we now invoke Eqs.(\ref{bigeasy}, \ref{equil}), we have
the following constraint on the bare spin current Eq.(\ref{js})
in equilibrium: 
\begin{eqnarray}
\langle {\bf j} \rangle = - \sum_{i}
\langle  \tau^{z}_{i} {\bf x}_{i}  \rangle.
\label{central_js}
\end{eqnarray}
This is so due to Eq.(\ref{equil}),
and the relation bewteen Eq.(\ref{shi}) and 
Eq.(\ref{js}). Similarly, the  conserved current ${\bf \overline{J}}$ may have a non-vanishing expectation value even at equilibrium:
\begin{eqnarray}
\langle {\bf \overline{J}} \rangle = - 2\sum_{i}
\langle \tau^{z}_{i} {\bf x}_{i}   \rangle+\ldots,
\label{central_joverline}
\end{eqnarray}
where, the omitted terms are higher order in the spin-orbit coupling.  When spin is a good quantum number, all three forms of the spin current vanish in equilibrium.  Only in the presence of spin-orbit coupling do the equilibrium currents differ.

\section{A variational proof using the 
Lieb-Schulz-Mattis twist operator}  
\label{A+}
We now regress and derive 
a version of
Eq.(\ref{bigeasy}) for  the ground state sector by a
generalization of the variational arguments of Ref.~\onlinecite{bohm}.  This
proof makes contact with topological Lieb-Schulz-Mattis \cite{lsm} type
twist operators. 

We consider the ground state
$|0 \rangle$ of a general system with spin-orbit coupling and prove by a variational argument that within it the spin current vanishes:
\begin{eqnarray}
\langle 0 | {\bf{J}} | 0 \rangle = 0,
\label{j0}
\end{eqnarray}
where, we take spin current $\bf{J}$ as in Eq.(\ref{shi}).  By definition, the ground-state energy 
\begin{eqnarray}
E_{0} \equiv \langle 0 | H |0 \rangle.
\label{e0}
\end{eqnarray}
is the lowest attainable
energy and lies at
the bottom of the spectrum. 
Let us assume that we have a state $|\psi\rangle$
with a finite (i.e. non-zero) spin current. 
We will then prove that $|\psi\rangle$ cannot be the ground-state, by constructing another state
$| \psi(\delta {\bf P})\rangle$ with a lower energy.  That is, if $\langle \psi | {\bf{J}} | \psi \rangle\neq 0$
then, the energy $E(\delta {\bf P})$ 
of the ``twisted" state $| \psi (\delta{\bf P}) \rangle$ can be made
to be lower than the energy $E(0)$ of $|\psi\rangle$, by choosing the variational parameter $\delta {\bf P}$ appropriately:
\begin{eqnarray}
E(\delta {\bf P}) < E(0).
\label{10}
\end{eqnarray}
In this way we establish that the ground-state $|0\rangle$ cannot carry a spin current of the form in Eq.~(\ref{shi}), since, by definition, it has the lowest energy.

We construct the variational state $|\psi(\delta{\bf P}) \rangle$ by applying a spin-dependent boost to the state $|\psi\rangle$: 
\begin{eqnarray}
\ket{\psi}&=&
\exp(- i \delta {\bf P}\cdot {\bf L})
\ket{\psi} \nonumber\\  
&=& \exp(- i \delta {\bf P}\cdot \sum_{i} {\bf x}_{i} \sigma^{z}_{i}),
\label{newstate}
\end{eqnarray}  
where, we have used the spin-displacement operator ${\bf L}$ from Eq.~(\ref{aspin}).  Compared to $\ket{\psi}$, the momentum of each particle in $\ket{\psi(\delta{\bf P})}$ with spin up $\sigma^z_i=+S$ changes by $+S\delta {\bf P} $ while the momentum of particles with spin down $\sigma^z_i=-1$ changes by $-S\delta {\bf P}.$  The momentum boost $\delta{\bf P}$ will be our variational parameter.  Lieb-Shulz-Mattis perform a similar construction but restrict the system geometry to be quasi-one-dimensional, and the spin $S$ to be half-integer.\cite{lsm}  We require no such restrictions: the conclusion Eq.~(\ref{j0}) holds in one-, two- and three-dimensions, for systems with half-integer spin particles as well as those containing integer spin particles.

The energy of $\ket{\psi(\delta {\bf P})}$ for small $\delta P$ may be expanded
\begin{eqnarray}
\bra{\psi(\delta {\bf P})}H\ket{\psi(\delta {\bf P})}=\bra{\psi}H\ket{\psi} &-& i\delta{\bf P}
\cdot \bra{\psi}[H,\sum_i \sigma^z_i \bf{x}_i] \ket{\psi}\nonumber\\ &+& {\cal O}((\delta P)^2).
\label{energy}
\end{eqnarray}
Now invoking the equation of motion $\langle d{\bf L}/dt \rangle = i \langle [H,{\bf L}]\rangle$ and the definition of the spin current ${\bf J} = d {\bf L}/dt$ in Eq.~(\ref{shi}) we have 
\begin{eqnarray}
E(\delta {\bf P}) = E(0) + \delta {\bf P}\cdot \langle \psi | {\bf J} | \psi \rangle + {\cal O}((\delta  P)^2).
\label{10final}
\end{eqnarray}
It follows that if the state $\ket{\psi}$ carries a spin current
$\langle \psi |{\bf J}|\psi \rangle \neq 0$
then there always exists a sufficiently small $\delta  P$
such that $E(\delta {\bf P})$ can be made smaller than $E(0)$, by choosing the direction of the momentum boost for spin-up particles $\delta {\bf P}$ opposite to the current direction $\langle {\bf J} \rangle$.  Since the ground-state has the lowest energy, we must have Eq.~(\ref{j0}), a special
case of the more general Eq.(\ref{bigeasy}).

\section{Conclusions and Discussion}
\label{D}
In spin-orbit coupled systems, the definition of spin current has inherent ambiguity.\cite{sonin2}  By redefining the current, it is possible to change its properties.  We consider three definitions of the spin current that generically differ with respect to the two properties that interest us most, continuity and persistence.    The bare spin current $\bf{j}$ defined in Eq.~(\ref{js}) may persist in equilibrium, but generally does not lead to spin accumulation.  The spin current $\bf{J}$ defined in ref.~\onlinecite{niu} obeys the conservation law (\ref{thatsit}) when the average torque vanishes.  However, the expectation $\langle n| {\bf J} | n\rangle$ vanishes in all energy eigenstates $|n\rangle$ of a system with a time-independent Hamiltonian.   A third definition (\ref{spdensity}) of the spin current $\overline{\bf{J}}$, obeys a continuity equation with a modified spin density $\overline{\rho}$, and may flow persistently in equilibrium.  

Expanding in powers of the spin-orbit coupling, we find that $\overline{\bf{J}}$ differs at leading order
from both the bare spin current $\bf{j}$ and the conserved current $\bf{J}$ of Ref.~\onlinecite{niu}.  This means that in principle, the expectation value $\langle n|\overline{\bf{J}}|n\rangle$ need not vanish, and, a conserved spin current $\overline{\bf{J}}$ may flow persistently even in equilibrium despite spin-orbit coupling.  We conclude that in general spin-orbit coupled systems, no definite relation exists among the concepts of persistence and conservation of spin current.  

We give a physical argument for the instability of persistent spin current of the form $\bf{J}$ defined in ref.~\onlinecite{niu}.  The argument makes contact with the Bloch's famous result for the instability of persistent charge currents.  In passing, we note a fundamental relation, summarized in (\ref{newstate}), between ${\bf J}$ the spin current of Ref.~\onlinecite{niu} and the twist generator used by Lieb, Schulz, and Mattis.  It would be interesting to formulate this relation in an $SU(2)$ invariant fashion.  

Several experiments bear directly on these issues.\cite{red,molenkamp,hasan}  Transport measurements in HgTe quantum wells provide evidence for gapless edge modes that are protected from non-magnetic impurity scattering and that respond singularly to an external magnetic field.\cite{molenkamp}  This is consistent with a ``helical'' picture for the edge modes, with a bare spin current $\bf{j}$ accompanying the charge current.  Spectroscopic measurements of the surface excitations in  Bi$_{1-x}$Sb$_x$ find an odd number of gapless Dirac points on the surface, suggesting topological stability of the gapless modes to non-magnetic impurities.\cite{hasan}   Further, in GaAs/GaAlAs quantum wells, time-resolved Kerr rotation experiments see evidence for a long-lived excitation mode that transports both spin and charge.\cite{red}

These systems contain strong spin-orbit coupling, so the bare spin current $\bf{j}$ accompanying the gapless charge modes need not give rise to a spin accumulation.  A detectable change in spin density $\rho$, could arise in response to an applied bias due to the flow of the current $\bf{J}$, provided that the average torque density vanishes.  For pure Rashba spin-orbit coupling, non-magnetic impurity scattering eliminates spin accumulation even under an applied bias.\cite{Mis}  The sytems described above are stable to non-magnetic impurities, opening up the possibility of detectable spin accumulation.

 
 We gratefully acknowledge discussions with A. V. Balatsky.
 ZN thanks the CMI of WU for support. NBA is grateful
 to Los Alamos National Laboratory and the National Center for Theoretical Sciences for their hospitality.

\begin{appendix}

\section{A derivation of conserved
spin current $\overline{\bf{J}}$ }
\label{cons}

We begin by considering the conservation of charge current. In the absence of spin-orbit coupling, this case also describes spin.  Charge conservation and gauge invariance are tightly linked. To see this in the simplest context, we consider a system with approximate $d+1$-dimensional Lorentz invariance such as graphene ($d=2$) or bismuth ($d=3$), and adopt the relativistic notation $A_\mu$, where, $A_0=\phi$ is the scalar potential and $A_a$ for $a=1,2,\ldots d$ are the components of the vector potential.  For convenience, we adopt the units $\hbar = c= 1$.  

Let us now suppose that the low-energy effective action $S_{eff}[A_\mu]$ is invariant under a transformation of the gauge field   
\begin{eqnarray}
A_{\mu} \to A_{\mu}+\delta A_{\mu}=A_{\mu} - \partial_{\mu} \Lambda
\label{gA}
\end{eqnarray}
with $\Lambda(x_\mu)$ any smooth function.  To see how this constrains charge flow, we formally expand the low-energy effective action in powers of the gauge field $A_\mu$,
\begin{equation}
S_{eff}[A_\mu]=\int d^dxdt {J_c}^{\mu} A_{\mu} + \ldots,
\label{effective}
\end{equation}
where, the charge current is given formally by the functional derivative $J_c^\mu=\frac{\delta S_{eff}[A_\mu]}{\delta A_\mu}$.  By definition of the charge current, the gauge transformation (\ref{gA}) leads to the following change in the effective action:  
\begin{eqnarray}
\delta S_{eff}&=& \int d^dxdt {J_c}^{\mu} \delta A_{\mu} \nonumber \\
&=& \int d^dxdt (\partial_\mu J_c^\mu )\Lambda + \ldots,
\end{eqnarray}
where, we have integrated by parts in the second line and dropped the contribution from the surface of the system.   Gauge invariance requires $\delta S_{eff}=0$ for arbitrary, smooth $\Lambda$, which can only be satisfied if the charge current is conserved:
\begin{eqnarray}
\partial_{\mu} J_c^{\mu} =0.
\label{cJ}
\end{eqnarray}
Conversely, charge conservation (\ref{cJ}) implies invariance of the effective action under the gauge transformation (\ref{gA}).  Thus, charge conservation and gauge invariance have an intimate relationship. 

To make contact with our results in Section II, let us express this connection between gauge invariance and charge conservation in the language of operators.  For a system
with time independent charges ($e_{i}$) the charge density operator is 
given by
\begin{eqnarray}
\rho_c = \sum_{i} e_{i} n_{i},
\label{cdensity}
\end{eqnarray}
with $n_{i}$ the number density operator.
If we now assume gauge invariance and the usual Maxwell dynamics for the gauge field, then we find charge is conserved at the operator level,
\begin{eqnarray}
\frac{\partial \rho_c}{\partial t}  
+ {\bf \nabla} \cdot {\bf J_c} =0,
\label{charge}
\end{eqnarray}
with the charge current operator given as follows:
\begin{eqnarray}
{\bf J_c} = \sum_{i} e_{i} {\bf v}_{i}. 
\label{jf}
\end{eqnarray}
Here ${\bf{v}}_{i}$ is the velocity of the $i$-th charge carrier in the system. 

For example, consider a single-particle with mass $m$ 
and charge $e$ in state $|\psi \rangle$.  Symmetrizing (\ref{jf}), we find the following familiar expression for the charge current (Here $a=1,2,\ldots d$ run over the spatial components): 
\begin{eqnarray}
\langle \psi| J_c^a | \psi \rangle = e  
\int d^{d}x \Big ( \psi^{*} [ \frac{1}{2mi} 
D_{a}] \psi - \psi [ \frac{1}{2mi} 
D_{a}] \psi^{*} \Big),
\label{London}
\end{eqnarray}
with the spatial components of the covariant derivative given by, $D_{a} = \frac{\partial}{\partial x^{a}} -
i e A_{a}$.  The velocity operator is given by ${\bf v}={\bf D}/m$, and we see here that the operator expression (\ref{jf}) describes the usual charge current in (\ref{London}).  

Let us now apply this to spin systems. As the spin states
change with time, we obviously cannot simply substitute
$e_{i} \to \sigma_{i}^{z}(t)$ in order to get
a conserved current. If we set, however, $e_{i} \to \sigma^{z}_{i}(t=0)$
then the steps above can
be reproduced word for word. 
Although, we will refer to spin currents
in what follows, our results hold for any
other ``charge'' field $\sigma$ which is allowed
to vary in time.  This is so as 
$\sigma^{z}(0)$-
the value of the $z$ component of the 
spin at time $t=0$- is, by definition, time independent. 
We may relate $\sigma^{z}(0)$ to $\sigma^{z}(t)$ 
by a time reversed evolution,
\begin{eqnarray}
\sigma^{z}(0) = e^{- i \int_{0}^{t} H(t') dt'} \sigma^{z}(t)
e^{i \int_{0}^{t} H(t') dt'}.
\end{eqnarray}
Here, $e^{i \int_{0}^{t} H(t') dt'}$ is a time ordered 
exponential. The total number operator $n_{i}$ (the sum 
over both up and down spin flavors) and its associated
number current satisfy the continuity equation (\ref{jf}), so the continuity equation for the spin current (\ref{spdensity}) given by (\ref{overline_conserve}) follows. 

It is interesting to formally expand the spin density $\overline{\rho}$ in powers of the spin-orbit coupling:
\begin{eqnarray}
\overline{\rho} = \sum_{i} \sigma_{i}^{z}(t=0) n_{i} \nonumber
\\ = \sum_{i}  \sigma_{i}^{z}(t) n_{i} - 
t \sum_{i} \frac{d \sigma_{i}^{z}}{dt} n_{i} \nonumber
\\ + \frac{t^2}{2} \sum_{i} \frac{d^{2} \sigma_{i}^{z}}{dt^{2}}  
n_{i} -\frac{t^3}{3!} 
\sum_{i} \frac{d^{3} \sigma_{i}^{z}}{dt^{3}}  
n_{i} + ...,
\end{eqnarray}
To zeroth order, $\overline{\rho}$ coincides with the bare spin density $\rho$.  However, already at leading order the two densities differ.

Turning now to the spin current, we can use the expression in (\ref{jf}) to write down a corresponding conserved spin current.  Again it is interesting to formally expand the expression in powers of the spin-orbit coupling:
\begin{eqnarray}
\overline{{\bf{J}} }= \sum_{i} \sigma_{i}^{z}(t=0) {\bf{v}}_{i} 
 \nonumber
\\ = \sum_{i} e^{-iHt} \sigma_{i}^{z}(t) e^{iHt} 
{\bf{v}}_{i}  \nonumber
\\ = \sum_{i} \sigma_{i}^{z} {\bf{v}}_{i}  
-  t \sum_{i} \frac{d \sigma_{i}}{dt} {\bf{v}}_{i}  \nonumber 
\\ + \frac{t^2}{2} \sum_{i} \frac{d^{2} \sigma_{i}^{z}}{dt^{2}}  
{\bf{v}}_{i}   - \frac{t^3}{3!} \sum_{i} \frac{d^{3} 
\sigma_{i}^{z}}{dt^{3}} {\bf{v}}_{i}  
+ \ldots .
\label{complete}
\end{eqnarray}
We find that to zeroth order in the spin-orbit coupling, $\overline{\bf{J}}$ coincides with the bare spin current $\bf{j}$ in (\ref{js}) and the conserved spin current $\bf{J}$ in (\ref{shi}).  At leading and higher order, however, the three currents differ. 


\end{appendix}

\end{document}